\documentclass[11pt]{article}
\usepackage{graphicx}
\usepackage{amssymb}
\usepackage{amsmath, amsthm}
\newcommand{\ee}{\end{equation}}
\newcommand{\word}[1]{\,\,\mbox{#1}\,\,}
\newcommand{\reff}[1]{(\ref{#1})}
\newcommand{\beq}{\begin{equation}}
\newcommand{\eeq}[1]{\label{#1}\end{equation}}
\newcommand{\beg}{\begin{equation*}}
\newcommand{\eeg}{\end{equation*}}

\newcommand{\eq}{\!=\!}
\newcommand{\p}{\!+\!}

\newcommand{\bsplit}{\begin{split}}
\newcommand{\esplit}{\end{split}}

\begin{document}
\def\theequation{\arabic{section}.\arabic{equation}}
\begin{titlepage}
\title{Compact dimensions and the Casimir effect:\\ the Proca connection}
\author{$^{1,2}$Ariel Edery\,\thanks{Email: aedery@ubishops.ca}\\$^1${\small\it Physics Department, Bishop's University}\\
{\small\it 2600 College Street, Sherbrooke, Qu\'{e}bec, Canada
J1M~0C8} \\\\ $^2$\thanks{work partly completed at KITP, Santa
Barbara} {\small\it Kavli Institute for Theoretical Physics,
 University of California}\\{\small\it Kohn Hall, Santa Barbara,
CA 93106 USA }
\\\\$^{3,4}$Valery N. Marachevsky\,\thanks{Email:
maraval@mail.ru}\\$^3${\small\it Laboratoire Kastler Brossel, CNRS,
ENS, UPMC,}\\{\small\it  Campus Jussieu case 74, 75252 Paris,
France}\\\\$^4${\small\it V. A Fock Institute of Physics, St.
Petersburg State University}\\{\small\it 198504 St. Petersburg,
Russia}}

\date{} \maketitle
\begin{abstract}
We study the Casimir effect in the presence of an extra dimension
compactified on a circle of radius $R$ ($M^4\times S^1$ spacetime).
Our starting point is the Kaluza Klein decomposition of the $5D$
Maxwell action into a massless sector containing the $4D$ Maxwell
action and an extra massless scalar field and a Proca sector
containing $4D$ gauge fields with masses $m_n=n/R$ where $n$ is a
positive integer. An important point is that, in the presence of
perfectly conducting parallel plates, the three degrees of freedom
do not yield three discrete (non-penetrating) modes but two discrete
modes and one continuum (penetrating) mode. The massless sector
reproduces Casimir's original result and the Proca sector yields the
corrections. The contribution from the Proca continuum mode is
obtained within the framework of Lifshitz theory for plane parallel
dielectrics whereas the discrete modes are calculated via $5D$
formulas for the piston geometry. An interesting manifestation of
the extra compact dimension is that the Casimir force between
perfectly conducting plates depends on the thicknesses of the slabs.

\end{abstract}
\setcounter{page}{1}
\end{titlepage}

\def\theequation{\arabic{section}.\arabic{equation}}

\section{Introduction}
\setcounter{page}{2} In the 1920's, Kaluza \cite{Kaluza} and Klein
\cite{Klein} attempted to unify classical gravitation and
electromagnetism by extending General Relativity to a
five-dimensional ($M^4\times S^1$) spacetime. In modern times,
string theory revived the idea of extra dimensions from a more
fundamental perspective. In this paper, we are interested in the
effect of a compact dimension on Casimir's original parallel plate
scenario. Before we discuss this subject, we note that the combined
effect of compact dimensions and Casimir energies has recently
attracted interest in cosmology \cite{Greene,Setare}. For example,
in a brane world scenario with toroidal extra dimensions, it was
found that Casimir energies can play a central role in attempting to
resolve some long outstanding puzzles. It was shown that under
certain conditions, Casimir energies can stabilize the extra
dimensions, allow three dimensions to grow large and provide an
effective dark energy in the large dimensions \cite{Greene}. Most
recently, there has been some interesting work on Casimir energies
in Randall Sundrum models \cite{Mariana-Frank,Cleaver} and in
one-dimensional piston scenarios with extra compact dimensions for a
massless scalar field obeying Dirichlet boundary conditions
\cite{Cheng}. The UV cut-off dependence of the Casimir energy for
perfect conductors with an extra compact dimension was also studied
recently in \cite{Leandros} (the compactification scheme is not
explicitly stated). Closer to the spirit of our work, the perfectly
conducting parallel plate scenario with an extra dimension
compactified to an $S^1/Z_2$ orbifold was studied in \cite{Bleicher}
and explicit analytic expressions for the Casimir force were
obtained.

In this paper, we calculate the correction to the Casimir force due
to an extra dimension compactified on a circle of radius $R$ for the
case of perfectly conducting parallel plates separated by a distance
$a$. Our starting point in Sec. $2$ is the Kaluza Klein (KK)
decomposition of the $4\p1$ ($5D$) Maxwell action into two sectors
in $4D$: a massless and massive (Proca) sector \cite{Raman}. The
massless sector contains the $4D$ Maxwell action as well as a $4D$
massless scalar field. The Proca sector yields an infinite set of 4D
massive gauge fields $A_{\mu}^{(n)}$ (and $A_{\mu}^{(n)*}\,$) with
masses $m_n =n/R$ where $n$ is a positive integer and $R$ is the
radius of the compact dimension. An advantage of the $KK$
decomposition is that the problem can now be analyzed in four
spacetime dimensions without reference to the extra fifth compact
dimension. In $5D$, the photon has three degrees of freedom or
polarizations and the $KK$ decomposition must yield the same number
of degrees of freedom since it describes the same physical system.
In the massless sector, the $4D$ Maxwell term yields the usual two
polarizations and the $4D$ massless scalar field yields one degree
of freedom for a total of three. In the $4D$ Proca sector, it is
well known that the massive photon has three polarizations because
of the presence of a longitudinal mode in addition to the usual two
transverse modes. Both sectors have three degrees of freedom as in
the original $5D$ case.

In Sec. $3$ the mode decomposition is explained in detail. An
important point is that in the presence of perfectly conducting
plates, not all three modes reflect perfectly at the boundary and
one does not obtain three discrete (non-penetrating) modes. The
three polarizations yield two discrete and one continuum
(penetrating) mode. This point does not seem to have been taken into
account in previous work \cite{Leandros,Bleicher}. In $4D$ Proca
theory, unlike the usual Maxwell theory in $4D$, there exists a
propagating continuum mode inside the perfect conductors
\cite{Dombey}. The electric and magnetic fields inside the perfect
conductors are zero but the gauge potentials are non-zero
contributing a non-zero energy density given by $m^2\,((A^{0})^2 \p
{\bf A}^2\,)$ where $m$ is the mass of the photon. The penetrating
continuum mode requires analyzing only the third component $A_z$ of
the gauge potential \cite{Dombey}. The boundary conditions on $A_z$
and its derivative $\partial_z\,A_z$ are that they be continuous at
the boundaries and this is equivalent to the boundary conditions on
a transverse-electric (TE) mode propagating in plane parallel
dielectrics of different permittivities. The Casimir contribution
from the continuum mode can therefore be calculated efficiently by
making use of Lifshitz theory \cite{Lifshitz}.

The massless sector reproduces exactly Casimir's original result
\cite{Casimir}. The Proca sector is responsible for the corrections
due to the extra compact dimension and these are derived in Sec.
$4$. The corrections from the discrete modes are a function of the
radius $R$ of the compact dimension and the correction from the
continuum mode is a function of both the radius $R$ and the
thicknesses $\ell_1$ and $\ell_2$ of the two perfectly conducting
slabs respectively. In the limit as $R\to 0$ both the Proca discrete
and continuum mode corrections vanish and one recovers Casimir's
original result.

Casimir force calculations for the Proca discrete and the Proca
continuum modes are fundamentally different. As already mentioned,
the continuum modes are calculated via Lifshitz theory. For the
discrete modes we derive in the appendix Casimir piston formulas for
Dirichlet boundary conditions that are a $5D$ generalization of
previously derived $4D$ expressions \cite{Marachevsky1}. We then
make use of the parallel plate limit of these formulas. The
advantage of the piston scenario \cite{Cavalcanti} is that it
automatically includes the Casimir contribution from the exterior
region. The piston separates the volume into an interior and
exterior and the main point is that contributions from both chambers
must be included in any realistic calculation of the Casimir force.
Some of the first exact results in $3\p1$ dimensions include the
$3\p 1$ Dirichlet piston \cite{Ariel1} and the $3\p1$ EM piston
\cite{Marachevsky1, Hertzberg} (see also
refs.\cite{ArielVal}-\cite{Lim2} with a review in \cite{ArielVal}).

We plot the contribution from both the Proca discrete and continuum
modes as a function of the circumference $d\eq 2\,\pi R$. The
discrete modes make a $12.1 \%$ correction to Casimir's original
result when $d\eq2$ (lengths are in units of the plate separation
$a$). The continuum mode yields a correction of $0.6\%$ at $d\eq2$
when the thickness of each slab is $\ell_1=10$ (this is a maximum at
$d\eq 2$ since $\ell_1=10$ yields results that are almost identical
to $\ell_1=\infty$). The total correction to Casimir's result at
$d\eq2$ and $\ell_1=10$ is therefore $12.7\%$ (the maximum
correction at $d\eq 2$). Both corrections decrease exponentially
fast as $d$ decreases reaching less than $1\%$ at $d\eq1$.

In Sec. $5$ we summarize our results and discuss some relevant
high-precision Casimir experiments that may be important in the
future to detect the effect of extra dimensions.

We use units with $\hbar=c=1$ throughout.

\section{Kaluza Klein decomposition of the $5D$ Maxwell action}
\setcounter{equation}{0}

This section follows closely the TASI lectures, ``To the Fifth
dimension and Back" by Raman Sundrum \cite{Raman} (the signature is
$(+,-,-,-,-)$). We will therefore be brief but explain enough for
the work to be self-contained and applicable to our particular case
(i.e. abelian gauge fields). The Maxwell action in $5D$ is given by
\begin{align} S&=\int d^4x \int dx^4 \Big\{-\dfrac{1}{4}
\,F_{a\,b}F^{a\,b}\Big\}\\&=\int d^4x \int dx^4
\Big\{-\dfrac{1}{4}\,F_{\mu\nu}\,F^{\mu\nu}-\dfrac{1}{2}\,F_{\mu\,4}\,F^{\mu\,4}\Big\}\label{action}\end{align}
where $a$ and $b$ are $5D$ indices (they run from $0$ to $4$
inclusively) and $\mu$ and $\nu$ are $4D$ indices (they run from $0$
to $3$ inclusively). If the fourth spatial dimension is compactified
to a circle of radius $R$ we can express $x^4$ as $R\,\phi$ where
$\phi$ is an angular coordinate $-\pi\le\phi\le\pi$. We then can
Fourier expand the (abelian) gauge fields as \beq A_b(x^{\mu},\phi)=
A_b^{(0)}(x) + \sum_{n=1}^{\infty} (A_b^{(n)}(x) \,e^{i\,n\,\phi}+
c.c.)\eeq{decomp} An important point is that it is not possible to
go to axial gauge $A_4\eq0$ because it is not possible to remove the
``n=0" part $A_4^{(0)}$. This would require a gauge transformation
with function $\Lambda=-x^4\,A_4^{(0)}$ so that $\partial_4 \Lambda=
-A_4^{(0)}$. However, such a $\Lambda$ is not valid because it is
proportional to $x^4$ and hence is not periodic. As pointed out in
\cite{Raman}, the closest to axial gauge one can reach is ``almost
axial" gauge where $A_4$ does not depend on $x^4$ i.e.
$A_4(x,\phi)=A_4^{(0)}(x)$. The action \reff{action} then becomes
\cite{Raman}
\begin{align} S&=\int d^4x \int_{-\pi}^{\pi} R\,d\phi
\Big\{-\dfrac{1}{4}\,F_{\mu\nu}\,F^{\mu\nu}
+\dfrac{1}{2}\,(\partial_{\mu}A_4^{(0)})^2
+\dfrac{1}{2}\,(\partial_4 A_{\mu})^2 \Big\} \\&= 2\,\pi\,R \int
d^4x \,\, \Bigg\{-\dfrac{1}{4} F_{\mu\nu}^{(0)}F^{\mu\nu(0)}
+\dfrac{1}{2}
(\partial_{\mu}\,A_4^{(0)})^2\\&\quad\quad+\,\sum_{n=1}^{\infty}\Big[
-\dfrac{1}{2}|\partial_{\mu}\,A_{\nu}^{(n)}-\partial_{\nu}\,A_{\mu}^{(n)}|^2
+\dfrac{n^2}{R^2} |A_{\mu}^{(n)}|^2 \Big]\Bigg\}\,.
\label{action2}\end{align} The 5D Maxwell action decomposes into a
$4D$ massless sector containing $4D$ Maxwell plus an extra scalar
field $A_4^0$ and a Proca sector with an infinite set of $4D$ gauge
fields $A_{\mu}^{(n)}$ (and $A_{\mu}^{(n)*}$) of mass $m_n=n/R$
where $R$ is the radius of the compact dimension and $n$ is a
positive integer.

\section{Mode decomposition in the presence of perfectly conducting
parallel plates}

\subsection{Massless Sector}
We now find the mode decomposition in the presence of perfectly
conducting parallel plates separated by a distance $a$ (situated at
$z=0$ and $z=a$). We begin with the massless sector and then look at
the Proca sector. In the massless sector, we have the usual four
gauge components $A_0^{(0)},A_1^{(0)},A_2^{(0)}$ and $A_3^{(0)}$ of
4D Maxwell plus an extra 4D massless scalar field $\phi\equiv
A_4^{(0)}$. This yields three degrees of freedom or polarizations:
the usual two polarizations from $4D$ Maxwell and one extra degree
of freedom from the massless scalar field $\phi$. We can go to
radiation gauge where $A_0^{(0)}\eq0$ and $\partial_i
A^{i^{(0)}}\eq0$ where $i=1,2,3$ (this can be extended to include
$i=4$ since $\partial_4\,A^{4^{(0)}}$ is identically zero). At the
surface of the plates the electric field components $E_1$ and $E_2$
are zero so that $A_1^{(0)}$ and $A_2^{(0)}$ are zero at the surface
of the conducting plates. The gauge condition $\partial_i
A^{i^{(0)}}\eq0$ yields a condition on $A_3^{(0)}$, namely
$\partial_3 \,A^{3^{(0)}}\eq0$ at the surface of the plates. The
field $\phi$ is a continuum mode that obeys the wave equation
$\square\,\phi=0$ for a free massless scalar field. The perfect
conductors do not impose any extra condition on the scalar field.
The mode decomposition for the massless sector is then given by
\begin{equation*}\left.\begin{aligned} A_0^{(0)}&=0
\\A_1^{(0)}&=c_1\, \sin(k_z\,z)\,e^{i({\bf k.x}-\omega t)}\,\\ A_2^{(0)}&=c_2\,
\sin(k_z\,z)\,e^{i({\bf k.x}-\omega t)}\,\\ A_3^{(0)}&=c_3\,
\cos(k_z\,z)\,e^{i({\bf k.x}-\omega t)}\,\\\phi=A_4^{(0)}&=c_4\,
e^{i({\bf k.x}+ p_z\,z-\omega' t)} \end{aligned}\right\}
\begin{aligned} k_z&=\dfrac{n\,\pi}{a}\quad n=1,2,3...\\{\bf
k}&=(k_x,k_y)\\{\bf x} &=(x,y)\\\omega^2&= {\bf k}^2 +k_z^2={\bf
k}^2 +\dfrac{n^2\,\pi^2}{a^2}\end{aligned}
\end{equation*}
where the momenta $p_z$ and $k_x, k_y$ are continuous. The three
modes $A_1^{(0)},A_2^{(0)}$ and $A_3^{(0)}$ yield two independent
modes because of the gauge condition $\partial_i\,A^i\eq0$. These
two modes are discrete containing the quantized momentum $k_z\eq
n\,\pi/a$. The third independent mode is the massless scalar field
$\phi$ and it is a continuum mode. The important point is that there
are three independent modes but only two are discrete. The scalar
field $\phi$ makes no contribution to the Casimir force between the
plates because it is a free field throughout the spacetime: it is
not influenced by the perfectly conducting boundary conditions
imposed on the electromagnetic field in $4D$. The two discrete modes
and their frequencies $\omega$ are identical to what appears in the
usual 4D case with perfectly conducting parallel plates. They
therefore reproduce exactly Casimir's result and we can simply quote
the result. The Casimir force per unit area (the pressure $P_0$)
stemming from the massless sector is given by \cite{Casimir} \beq
P_0= -\dfrac{\pi^2}{240\,a^4}\,.\eeq{ds}

\subsection{Proca sector: two discrete modes and one continuum mode}

The Casimir effect for photons of mass $m$ (Proca theory) in the
presence of perfectly conducting plates was analyzed in detail by
Barton and Dombey \cite{Dombey}.  In the presence of perfectly
conducting parallel plates, Barton and Dombey showed that the three
polarizations in Proca theory yield two discrete modes and one
continuum mode. The important point is that the three polarizations
do not yield three but two discrete modes. The discrete modes are
perfectly reflected or non-penetrating modes where the gauge fields
$A_{\mu}$ are zero both inside and on the surface of the conductor
\cite{Dombey}. The latter condition (i.e. gauge potentials are zero
on the surface) stems from the continuity of the gauge potentials in
Proca theory. The contribution of the two discrete modes does not
depend on the thickness of the conductors and yields an
$R$-dependent correction to Casimir's original parallel plate result
$P_0$ given by \reff{ds}. The continuum mode penetrates through the
conductors and makes a contribution to the Casimir force that
depends on the thicknesses $\ell_1$ and $\ell_2$ of the two
conducting slabs and the radius $R$ of the compact dimension. This
contribution enters into the Casimir force in a fundamentally
different way than the discrete modes. Both the continuum and
discrete mode corrections vanish in the limit $R \to 0$.

\subsubsection{Discrete modes}
With the Lorentz condition $\partial^{\mu}\,A_{\mu}^{(n)}=0$, the
equations of motion outside the conductor are given by $(\square +
m_n^2 \,)\,A_{\mu}^{(n)}=0$ where $m_n^2=n^2/R^2$. For the discrete
modes, the perfect conductor boundary conditions are that the gauge
components $A_0^{(n)}, A_1^{(n)},A_2^{(n)}$ and $A_3^{(n)}$ are zero
on the surface of the conductor (at $z=0$ and $z=a$).
 The mode decomposition for the discrete
modes is given by \cite{Dombey}
\begin{equation*}\left.\begin{aligned} A_0^{(n)}&=c_0\, \sin(k_z\,z)\,e^{i({\bf k.x}-\omega t)}\,
\\A_1^{(n)}&=c_1\, \sin(k_z\,z)\,e^{i({\bf k.x}-\omega t)}\,\\ A_2^{(n)}&=c_2\,
\sin(k_z\,z)\,e^{i({\bf k.x}-\omega t)}\,\\ A_3^{(n)}&=0
\end{aligned}\right\} \begin{aligned}k_z&=\dfrac{\ell\,\pi}{a}\quad(\ell=1,2,..)\\{\bf
k}&=(k_x,k_y)\;;\;{\bf x} =(x,y)\\\omega^2&= {\bf k}^2 +k_z^2 +
m_n^2
\end{aligned}\end{equation*} Note that $A_3^{(n)}$
cannot be a discrete mode (either a $\cos(k_z \,z)$ or $\sin(k_z
\,z)$ term) because $A_3^{(n)}$ must be both zero on the surface of
the conductor and satisfy the Lorentz condition. The three modes
$A_0^{(n)}, A_1^{(n)}$ and $A_2^{(n)}$ together with the Lorentz
condition yield two independent discrete modes for every $n$. These
have frequency $\omega$ given by \beq \omega^2= {\bf k}^2 +k_z^2 +
m_n^2 = k_x^2 +k_y^2 + \dfrac{\ell^2\,\pi^2}{a^2} +
\dfrac{n^2}{R^2}\,.\eeq{freq} where $\ell$ and $n$ are both positive
integers that run from $1$ to $\infty$. The same analysis applies to
the fields $A_{\mu}^{(n)*}$. The parallel plate geometry can be
thought of as a rectangular box with plate separation $a$ and plate
area $b\times c$ where $b$ and $c$ are taken to be large (infinite
limit). The continuous momenta $k_x$ and $k_y$ can be expressed as
$n_x \pi/b$ and $n_y\,\pi/c$ in the limit $b,c\to \infty$, with
$n_x$ and $n_y$ positive integers. The frequency is then given by
\beq \omega=
\dfrac{\ell^2\,\pi^2}{a^2}+\dfrac{n_x^2\,\pi^2}{b^2}+\dfrac{n_y^2\,\pi^2}{c^2}
 + \dfrac{n^2}{R^2} \word{where} b \word{and} c \word{are assumed large (infinite limit)}\,.\eeq{freq}
Let $E_D$ be the $4+1$ dimensional Dirichlet Casimir energy defined
by the sum over the four quantum numbers $\ell, n_x,n_y$ and $n$ of
$\omega/2$ from $1$ to $\infty$ (i.e. each of the four sums starts
at $1$). The Casimir energy for the Proca discrete modes $E_{prd}$
is then equal to four times the Dirichlet Casimir energy $E_D$:\beq
E_{prd}= 4\,E_D \,.\eeq{proca} The factor of four stems from the
Proca sector having two discrete modes for $A_{\mu}^{(n)}$ and two
discrete modes for $A_{\mu}^{(n)*}$.

\subsubsection{Continuum mode and the equivalence with TE mode in dielectric}
\begin{figure}[ht]
\begin{center}
\includegraphics[scale=0.60]{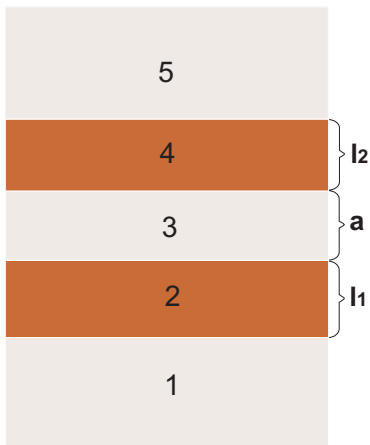}
\caption{The parallel plate set-up with its five regions. The five
regions starting with region 1 are: exterior (vacuum), conducting
slab of thickness $\ell_1$, vacuum gap of length $a$, conducting
slab of thickness $\ell_2$, exterior (vacuum).}
\end{center}
\end{figure}
Besides the two discrete modes, Proca theory in the presence of
conducting plates has a continuum (penetrating) mode which also
contributes to the Casimir energy. This has been analyzed in
\cite{Dombey}. In our case of an extra compact dimension we replace
the mass $m$ in Proca theory by the mass $m_n=n/R$ (we then sum over
all $n$ when calculating the Casimir energy).

The continuum mode requires analyzing only the component $A_z$
\cite{Dombey} (in our case we have $A_z^{(n)}$ for each $n$). In the
Proca theory vector potentials of the electromagnetic field are
continuous and satisfy the Lorentz condition. The latter implies
that $\partial_z A_z^{(n)}$ is continuous across the boundary since
$\partial_t A_0^{(n)}$ and $\nabla_{||} {\bf A}_{||}^{(n)}$ are
continuous across the boundary. Therefore in Proca theory the field
$A_z^{(n)}$ and its normal derivative $\partial_z A_z^{(n)}$ are
continuous across the flat boundaries perpendicular to the $z$
direction \cite{Dombey}.

The electric field is zero inside the perfect conductors. From this
condition and the Lorentz condition it follows that the $A_z^{(n)}$
components satisfy the equation for a massless scalar field inside
the conductors $\square A_z^{(n)}\eq0$. Outside the media (in
vacuum) the $A_z^{(n)}$ components satisfy the equation for a
massive scalar field $(\,\square + m_n^2\,)\, A_z^{(n)}\eq 0$. In
the parallel plate set-up there are five regions (see Fig.1). Region
1 is the exterior vacuum, region 2 is a conducting slab of thickness
$\ell_1$, region 3 is the gap of length $a$ (in vacuum), region 4 is
a conducting slab of thickness $\ell_2$ and region 5  is the
exterior vacuum. Let $m$ and $m=0$ correspond to massive and
massless propagation. From region 1 to region 5, we therefore have
the sequence $m,m=0,m,m=0,m$.

The equations of motion for $A_z^{(n)}$ and the conditions of
continuity for $A_z^{(n)}$ and its derivative $\partial_z A_z^{(n)}$
are exactly equivalent to the conditions for the transverse electric
(TE) electromagnetic mode propagating in slabs of different
dielectric permittivities. Massive propagation is equivalent to TE
propagation in a dielectric with permittivity
$\epsilon(\omega)=1-\omega_p^2/\omega^2$ where the plasma frequency
is given by $\omega_p=m_n$. Massless propagation is equivalent to
propagation in a dielectric with permittivity $\epsilon\eq1$. The
conditions of continuity for the TE mode at the boundaries are the
same as those on $A_z^{(n)}$ and its derivative. Therefore if the
five regions are replaced by dielectrics with permittivities
$\epsilon(\omega), \epsilon=1, \epsilon(\omega),
\epsilon=1,\epsilon(\omega)$ corresponding to the sequence
$m,m=0,m,m=0,m$ we obtain a physically equivalent set-up. The upshot
of all this is that the Casimir energy of the continuum mode
$A_z^{(n)}$ can be calculated in the framework of Lifshitz theory.

The same conclusions can be applied to the continuum modes
$A_z^{(n)*}$ and they make the same contribution to the Casimir
energy as $A_z^{(n)}$.

\section{Corrections to the Casimir force}

\subsection{Proca discrete modes}

The Dirichlet Casimir energy $E_D$ in a piston geometry associated
with the frequency $\omega$ is given by (see appendix) \beq E_D
=-\dfrac{1}{2\pi}\sum_{kmn=1}^{\infty}\sum_{\ell=1}^{\infty}\dfrac{\lambda_{kmn}}{\ell}\,K_1(2\,\ell\,\lambda_{kmn}\,a)
\word{with} \lambda_{kmn}=\sqrt{(\pi k/b)^2+(\pi m/c)^2+ (n/R)^2}
\eeq{Drich2} The sum $\sum_{kmn=1}^{\infty}$ is short-hand for a
triple sum with each sum running from $1$ to $\infty$. The
expression \reff{Drich2} for the Dirichlet energy in a piston
geometry automatically includes the contribution from both the
interior and exterior regions of the plates. The Casimir energy due
to Proca discrete modes is given by $E_{prd}=4\,E_D$. The Casimir
force due to Proca discrete modes is given by the negative
derivative with respect to the plate separation $a$:
\beq\begin{split}
F_{prd}&=-\dfrac{\partial}{\partial a}\,E_{prd}\\
&=-\dfrac{2}{\pi}\sum_{kmn=1}^{\infty}\sum_{\ell=1}^{\infty}\Big\{
\,\dfrac{\lambda_{kmn}}{a\,\ell}\,K_1(2\,\ell\,\lambda_{kmn}\,a)+2\,\lambda_{kmn}^2
\,K_0(2\,\ell\,\lambda_{kmn}\,a)\Big\}  \end{split} \,.\eeq{kj2} The
force per unit area is given by $F_{prd}/(b\,c)$ in the limit
$b,c\to\infty$. The sums over $k$ and $m$ in \reff{kj2} are
therefore replaced by integrals which can be expressed in terms of
modified Bessel functions:
\begin{flalign} \label{integrals}&\sum_{k=1}^{\infty}\sum_{m=1}^{\infty}
\lambda_{kmn}\,K_1(2\,\ell\lambda_{kmn}\,a)&\quad&\sum_{k=1}^{\infty}\sum_{m=1}^{\infty}
\lambda_{kmn}^2\,K_0(2\,\ell\lambda_{kmn}\,a)\nonumber\\
&\to\dfrac{b\,c}{2\,\pi}\int_0^{\infty}\!\!\!\!\sqrt{\smash[b]{r^2+\lambda_{n}^2}}\,\,K_1\big(\,2\,\ell\,a\,\sqrt{\smash[b]{r^2+\lambda_{n}^2}}\:\big)\,r\,dr
&\quad &\to \dfrac{b\,c}{2\,\pi}\int_0^{\infty}\!\!\!(r^2+\lambda_{n}^2)\,K_0\big(\,2\,\ell\,a\,\sqrt{\smash[b]{r^2+\lambda_{n}^2}}\:\big)\, r\,dr\nonumber\\
&=\Big(\dfrac{b\,c}{2\,\pi}\Big)\Big(\dfrac{1}{2\,\ell\,a}\Big)
\,\,\lambda_n^2\,K_2(2\,\ell\,a\,\lambda_n)\,.&\quad
&=\Big(\dfrac{b\,c}{4\,\pi}\Big)\Big(\dfrac{1}{\ell^2\,a^2}\Big)\,\,\lambda_{n}^{2}\,K_{2}(2\,\ell\,a\,\lambda_{n})
\nonumber\\&&&\qquad\qquad+\Big(\dfrac{b\,c}{4\,\pi}\Big)\Big(\dfrac{1}{\ell\,a}\Big)\,\,\lambda_{n}\,K_{1}(2\,\ell\,a\,\lambda_{n})\,.
\end{flalign}
After substituting the above in \reff{kj2} with $\lambda_n=n/R$, the
force per unit area due to the Proca discrete modes is equal to \beq
P_{prd}=\lim_{b,c\to\infty}\dfrac{F_{prd}}{b\,c}=-\sum_{n=1}^{\infty}
\sum_{\ell=1}^{\infty}\Big[\,\,\dfrac{3}{2}\dfrac{n^2}{\,\pi^2\,R^2\,a^2\,\ell^2}\,\,K_2(2\,\ell\,n\,a/R)
+\dfrac{n^3}{\ell\,\pi^2\,R^3\,a}\,K_1(2\,\ell\,n\,a/R)\Big]\,.
\eeq{Pcorr1} The correction coming from the Proca discrete modes in
units of Casimir's parallel plate result $P_0$ is \beq
\dfrac{P_{prd}}{P_0}=\sum_{n=1}^{\infty}\sum_{\ell=1}^{\infty}\Big[\,\,\dfrac{360}{\pi^4}\dfrac{n^2}{\ell^2}\Big(\dfrac{a^2}{R^2}\Big)\,K_2(2\,\ell\,n\,a/R)
+\dfrac{240}{\pi^4}\dfrac{n^3}{\ell}\dfrac{a^3}{R^3}\,K_1(2\,\ell\,n\,a/R)\,\,\Big]\,.\eeq{parallel}
In Fig. 2, we plot \reff{parallel} as a function of the
circumference $d=2\,\pi\,R$ in units of the plate separation $a$.
The correction is $12.1\%$ of Casimir's result when $d=2$ (or
$R=1/\pi$) but decreases rapidly (exponentially) as $d$ decreases.
It is $2.7\%$ when $d=1.5$, $0.098\%$ when $d=1$ and $16 \times
10^{-7}\%$ when $d=0.5$.

When compact dimensions are {\it not} present, the piston geometry
does not modify Casimir's parallel plate result in the limit of
infinitely sized plates. In this limit the exterior contribution to
the Casimir energy is equal to a regularized volume term of an
exterior region, which is equal to zero in the zeta function
regularization or should be subtracted in other regularizations.
However, when compact dimensions are present, the exterior region
makes a non-zero contribution even for the case of infinitely sized
parallel plates. Therefore, to calculate correctly the effect of a
compact dimension on perfectly conducting parallel plates, one must
include not only the correction due to the interior but also the
correction due to the exterior in all regularizations. The
correction \reff{Pcorr1} includes a significant contribution from
the exterior. The relevance of the piston geometry is therefore
highlighted by the presence of compact dimensions. As an
illustration, for the case of $a=1$ and $R=1/\pi$ (or $d=2$),
\reff{Pcorr1} yields \beq
 P_{prd}= -0.004990022267 \quad\word{when} R=1/\pi.
\eeq{da1} To determine the interior and exterior contributions to
this result we can use formulas for the Dirichlet Casimir piston
force $F_D$ derived in \cite{Ariel3} and then evaluate
$F_{prd}=4\,F_D$.  The Dirichlet piston force $F_D$ includes a
contribution from the interior ($F_{D_I}$) and exterior
($F_{D_{II}}$). The interior and exterior contributions can be
obtained from equations (3.15) and (3.18) in \cite{Ariel3}
respectively. We do not explicitly write them out here but simply
evaluate them when $R=1/\pi$:
\begin{align} \lim_{b,c\to\infty} \dfrac{F_{D_I}}{b\,c}
&=\dfrac{\pi}{2^5}\Big(3\,\Gamma(2)\,\pi^{-3}\,\zeta(4)-
4\,\Gamma(5/2)\,\pi^{-7/2}\,\zeta(5)+8\sum_{n=1}^{\infty}\sum_{\ell=1}^{\infty}\dfrac{n^3}{\ell}\,
K_1(2\,\pi\,n\,\ell)\Big)\nonumber\\&=0.00121490257\,. \\\nonumber\\
\lim_{b,c\to\infty} \dfrac{F_{D_{II}}}{b\,c} &=-\dfrac{\pi}{2^5}
\Gamma(5/2)\,\pi^{-7/2}\,\zeta(5)=-0.00246240814\,.\end{align} We
therefore obtain \beq \lim_{b,c\to\infty} \dfrac{F_{prd}}{b\,c}
=4\,(0.00121490257-0.00246240814) =-0.004990022267\eeq{final} which
is in agreement with \reff{da1}.

The important point is that the exterior contribution $F_{D_{II}}$
is not negligible and in fact, for the case we considered has a
higher magnitude then the interior $F_{D_I}$. Moreover, note that
$F_{D_I}$ here is positive and it is only the total
$F_D=F_{D_I}+F_{D_{II}}$ which is negative (as it must be since
$F_{prd}$ given by \reff{kj2} is manifestly negative). Without the
exterior contribution one would erroneously conclude that there is a
repulsive force.

\subsection{Proca continuum mode}

As already explained, the Casimir energy due to the continuum mode
can be calculated via Lifshitz theory \cite{Lifshitz, Ginzburg,
Milton1, Brevik1}. We assume that the five regions in Fig. 1 are
filled with dielectrics of permittivities (starting with region 1)
$\epsilon(\omega), \epsilon=1, \epsilon(\omega),
\epsilon=1,\epsilon(\omega)$ where
$\epsilon(\omega)=1-\omega_p^2/\omega^2$ with the plasma frequency
given by $\omega_p=m$. The Casimir energy of a TE mode propagating
in the dielectrics is then equivalent to the Casimir energy of a
Proca continuum mode of mass $m$ (we will later replace $m$ by
$m_n=n/R$). The Casimir energy $E_c$ for the continuum is thus given
by ($S$ is the surface of the plates):
\begin{align}
E_c (l_1, a, l_2, m) = &S \int_0^{+\infty} \frac{d\omega}{2\pi}
\int_0^{+\infty} \frac{2\pi k dk }{(2\pi)^2} \, {\rm ln} f (i\omega,
k, l_1, a, l_2, m)\label{refl}
\end{align}
where
\begin{align}
f (i\omega, k, l_1, a, l_2, m) &=
1 - r_{down}(i\omega, k, l_1, m) \,r_{up}(i\omega, k, a, l_2, m) ,\\
r_{down}(i\omega, k, l_1, m) &= \frac{(\rho_2+\rho_1)(\rho_3 -
\rho_2)+ (\rho_2- \rho_1)(\rho_3+\rho_2)\,e^{-2\rho_2 l_1}
}{(\rho_2+\rho_1)(\rho_3+\rho_2)+
(\rho_2-\rho_1)(\rho_3-\rho_2)\,e^{-2\rho_2 l_1} } , \\
r_{up}(i\omega, k, a, l_2, m)
&=\frac{(\rho_4+\rho_5)(\rho_3-\rho_4)+(\rho_4-\rho_5)(\rho_3+\rho_4)
\,e^{-2\rho_4 l_2} }{(\rho_4 +\rho_5)(\rho_3+\rho_4) +
(\rho_4-\rho_5)(\rho_3-\rho_4)\,e^{-2\rho_4 l_2} }\, e^{-2 \rho_3 a}
\label{rup}
\end{align}
with definitions:
\begin{equation}
\rho_1^2 \eq \rho_3^3 \eq \rho_5^2 = k^2+\omega^2 +m^2 , \qquad
\rho_2^2\eq\rho_4^2 = k^2+\omega^2.
\end{equation}
Here $r_{down}\,(\omega, k, l_1, m) $ and $r_{up}\,(\omega, k, a,
l_2, m)$ are reflection coefficients of the downward and upward
going plane waves reflecting at the lower ($z=0$) and upper ($z=a$)
boundaries of the layer $3$ respectively (see \cite{gratings} for a
derivation of formulas analogous to (\ref{refl})). The factor $e^{-2
\rho_3 a}$ in (\ref{rup}) appears due to a mirror translation of the
upper boundary from a position $z=0$ to a position $z=a$ or, in
other words, after the change of coordinates $z=-z^{\prime} + a$.

It is convenient to switch to polar coordinates: $r^2\eq
k^2+\omega^2$ with $k\eq r\,\cos(\theta)$. Then \reff{refl} can be
expressed as \beq E_c=\frac{S}{(2\pi)^2} \int_0^{+\infty } r^2 \ln f
(r, l_1, a, l_2, m)\,dr.\eeq{fr} The $\rho$'s are now given by
\begin{equation}
\rho_1 \eq \rho_3 \eq \rho_5 = \sqrt{r^2 +m^2} , \qquad
\rho_2\eq\rho_4 \eq r
\end{equation}
and we obtain \beq f (r, l_1, a, l_2, m) = 1 - r_{down}(r, l_1, m)
\,r_{up}(r, a, l_2, m)\eeq{hg} where \begin{align}
r_{down}\,r_{up}&=
\dfrac{m^4\,(1-e^{-2r\ell_1})\,(1-e^{-2r\ell_2})\,e^{-2\sqrt{r^2+m^2}\,a}}{(r+\sqrt{r^2+m^2})^4-
m^4\,(e^{-2r\ell_1}+e^{-2r\ell_2}) +(\sqrt{r^2+m^2}-r)^4
\,e^{-2r(\ell_1+\ell_2)}}\,.\label{rdru}\end{align} We now replace
$m$ by $m_n=n/R$ in \reff{rdru} and also make the substitution
$u=r\,R$:
\begin{align}
r_{down}\,r_{up}&=g\,(n,u,\ell_1,\ell_2,R)\,\,e^{-2\,\sqrt{u^2+n^2}\,a/R}\label{rdru2}\end{align}
where \beq g = \dfrac{n^4\,(1-e^{-2u\ell_1/R})\,(1-e^{-2u\ell_2/R})}
{(u+\sqrt{u^2+n^2})^4- n^4\,(e^{-2u\ell_1/R}+e^{-2u\ell_2/R})
+(\sqrt{u^2+n^2}-u)^4 \,e^{-2u(\ell_1+\ell_2)/R}}\,. \eeq{grf} Note
that $g$ is independent of the plate separation $a$. The
contribution to the Casimir energy from  Proca continuum modes
$E_{prc}$ is given by
\begin{equation}
E_{prc}= 2 \sum_{n=1}^{\infty} E_c(l_1, a, l_2, n/R)
\end{equation}
where the factor of two includes the contributions of $A_z^{(n)}$
and $A_z^{(n)*}$. The Casimir pressure from the Proca continuum
modes is then given by
\begin{align} P_{prc}(l_1, a ,l_2, R)&=- \frac{2}{S} \sum_{n=1}^{\infty}
\frac{\partial E_c(l_1, a ,l_2, n/R)}{\partial a}
\\&=-\dfrac{1}{\pi^2\,R^4}\sum_{n=1}^{\infty}\int_0^{\infty}u^2\,\sqrt{u^2+n^2}
\dfrac{g\,e^{-2\sqrt{u^2+n^2}\,a/R}}{1-g\,e^{-2\sqrt{u^2+n^2}\,a/R}}\,du\,.
\end{align}
The above formula for the pressure due to the Proca continuum modes
converges exponentially fast and is dependent on the thicknesses
$\ell_1$ and $\ell_2$ of the perfect conductors. In Fig. 3,
$P_{prc}$ is plotted as a function of the circumference
$d\eq2\,\pi\,R$ and the thickness of the two slabs. In Fig. 3,
lengths are expressed in units of the plate separation $a$,
$P_{prc}$ is in units of Casimir's result $-\pi^2/(240\,a^4)$ and
the two slabs are assumed to have equal thickness
($\ell_1\eq\ell_2$). The pressure $P_{prc}$ increases as the
circumference $d$ of the compact dimension increases and as the
thickness $\ell_1$ of the conductors increases. In the limit $d\to
0$ ($R\to0$), the pressure $P_{prc}$ tends to zero exponentially
fast regardless of the thickness of the slabs. Conversely, in the
infinitely thin limit of the conductors where $\ell_1\to0$,
$P_{prc}$ tends to zero regardless of the value of $R$. For $d\eq
2$, $P_{prc}$ reaches a maximum of approximately $0.6\%$ of
Casimir's result (the maximum value is reached in the $\ell_1\to
\infty$ limit. This is close to the $\ell_1=10$ result in Fig. 3).
For a given $d$, the maximum value of $P_{prc}$  is significantly
less than the contribution $P_{prd}$ from the Proca discrete modes
(plotted in Fig. 2). For example, at $d\eq 2$, $P_{prd}$ makes a
$12.1\%$ contribution while $P_{prc}$ makes a maximum contribution
of only $0.6\%$. Hence, the bulk of the correction to Casimir's
result due to the presence of the compact dimension stems from the
Proca discrete modes.

The total Casimir pressure $P$ on the plates is obtained by summing
the contribution $P_0$ from the massless sector and the contribution
$P_{prd}+P_{prc}$ from the Proca sector:
\begin{align}P=&P_0 + P_{prd} + P_{prc} = \nonumber \\= &-
\dfrac{\pi^2}{240\, a^4}-\sum_{n=1}^{\infty}
\sum_{\ell=1}^{\infty}\Big[\,\,\dfrac{3}{2}\dfrac{n^2}{\,\pi^2\,R^2\,a^2\,\ell^2}\,\,K_2(2\,\ell\,n\,a/R)
+\dfrac{n^3}{\ell\,\pi^2\,R^3\,a}\,K_1(2\,\ell\,n\,a/R)\Big]  \nonumber \\
&-\dfrac{1}{\pi^2\,R^4}\sum_{n=1}^{\infty}\int_0^{\infty}u^2\,\sqrt{u^2+n^2}
\dfrac{g\,e^{-2\sqrt{u^2+n^2}\,a/R}}{1-g\,e^{-2\sqrt{u^2+n^2}\,a/R}}\,du
\label{PEM}\end{align} where $g$ is given by \reff{grf}. Equation
\reff{PEM} is our final result. Both $P_{prd}$ and $P_{prc}$ tend to
zero as $R\to 0$ and one recovers Casimir's parallel plate result
$P_0=- \tfrac{\pi^2}{240\, a^4}$ in this limit. Since $P_{prd}$ and
$P_{prc}$ are manifestly negative, the magnitude of the Casimir
pressure increases in the presence of the extra compact dimension
(it becomes more negative). Our final result \reff{PEM} naturally
differs from previous results \cite{Leandros, Bleicher} because the
correction to Casimir's result is not derived using three discrete
modes but two discrete and one continuum mode. Had we used three
instead of two discrete modes the correction from the Proca discrete
modes would have been a factor of $3/2$ times higher (e.g. $18.2\%$
instead of $12.1\%$ for a circumference $d\eq 2$). Instead of a
third discrete mode, we obtain a continuum mode which is
qualitatively and numerically different.

\section{Conclusions}

More than two decades ago, the Casimir effect in the $4D$ Proca case
where the photon has a mass $m$ was studied for perfectly conducting
parallel plates \cite{Dombey,Davies}. Barton and Dombey
\cite{Dombey} showed that in the presence of conductors there are
two discrete modes and one continuum mode and not three discrete
modes. Both the discrete and continuum modes contributed corrections
to Casimir's result. They derived the leading continuum mode
contribution for a small mass $m$ and found that it vanishes in the
limit $m\to 0$. Casimir's parallel plate result was recovered in the
limit $m\to0$.

The $5D$ Maxwell problem with one dimension compactified to a circle
is closely related to the Proca problem. As in the Proca case, the
polarizations in the presence of conducting plates yield discrete
and continuum modes. The existence of continuum modes which
propagate inside the perfect conductors is a qualitative distinction
between the $5D$ ($M^4\times S^1$) and the $4D$ Maxwell problem. The
Casimir force in the $5D$ spacetime depends on the thicknesses of
the slabs even for perfectly conducting boundary conditions, which
is an interesting manifestation of the extra compact dimension. We
derive exact results for the contributions of both the discrete and
continuum modes and our final result for the Casimir pressure on the
plates is given by \reff{PEM}. The corrections to Casimir's result
are manifestly negative and increase the magnitude of the Casimir
pressure on the plates. We plot the corrections versus the
circumference $d\eq2\pi R$ for the discrete and continuum modes. For
the case $d=2$ (in units of the plate separation) the correction is
$12.1\%$ for the discrete modes and a maximum of $0.6\%$ for the
continuum modes. Both contributions decrease exponentially fast as
$d$ decreases and Casimir's parallel plate result is recovered in
the limit as the radius $R$ of the compact dimension tends to zero.

The parallel plate geometry is a natural theoretical bench-mark for
calculating new effects such as those originating from extra compact
dimensions.  Casimir experiments involving parallel plates are
notoriously difficult and the most precise experiments to date have
reached only $15\%$ precision \cite{Onofrio}. Measurements have also
been carried out to a precision of $1\%$ for sphere-plate
separations in the range $0.1-0.9 \mu m$ in another set of
experiments \cite{Mohideen}. The experiments involving a
micromachined torsional oscillator now lead to a precision better
than $1\% $ \cite{Decca}. In these experiments one measures a
gradient of the Casimir force between a sphere and a plate. The
gradient of the force between the sphere and the plate can be
expressed for a wide range of distances in terms of the force
between the two plates $F_{pp}$: $F^{\prime}_{PS}= 2 \pi R_S F_{PP}
$, where $R_S$ is the radius of the sphere. Thus the theoretical
results for the Casimir force between the two plates can be verified
with a remarkable precision.

\begin{figure}[ht]
\begin{center}
\includegraphics[scale=0.70]{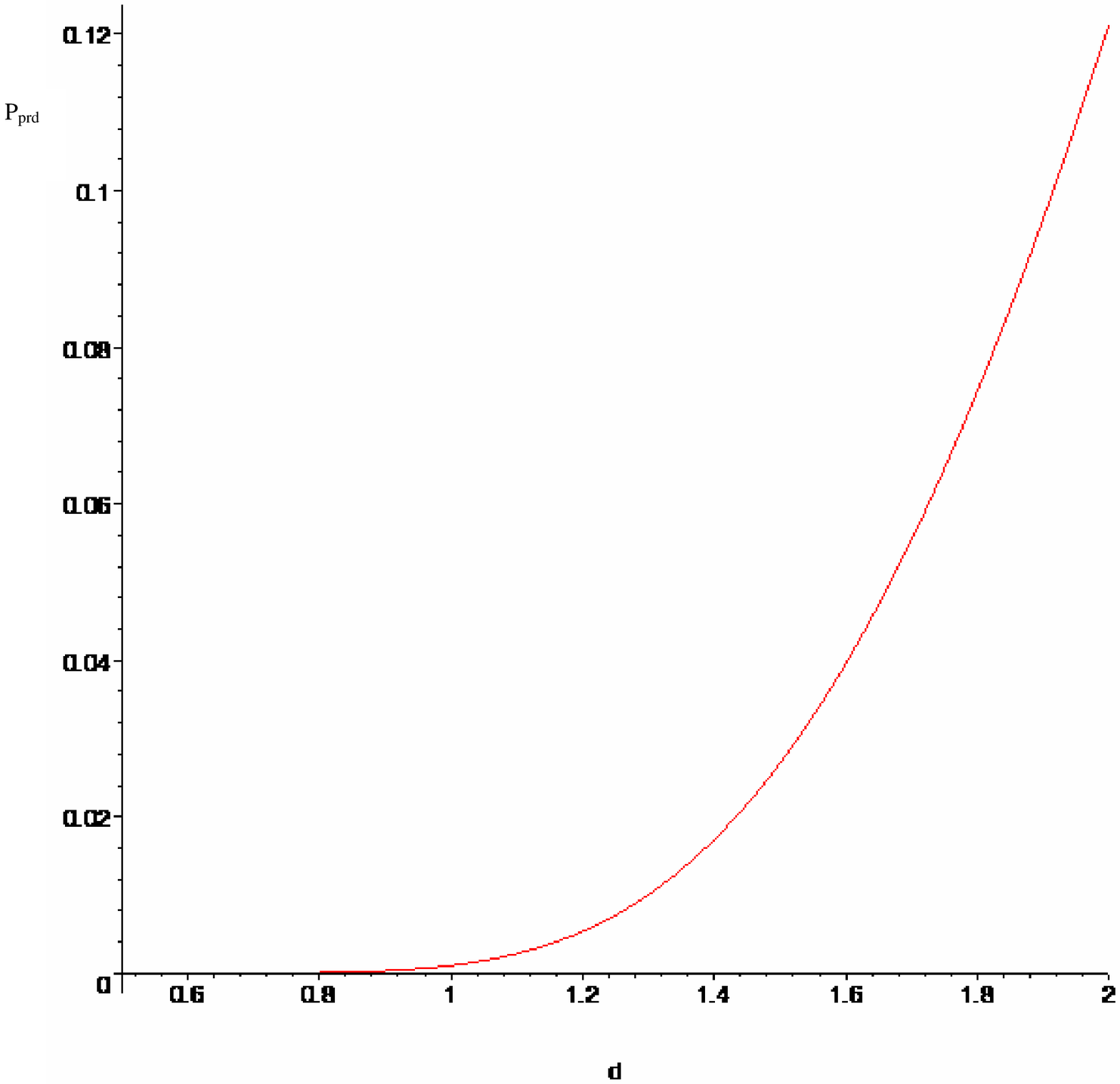}
\caption{Correction from Proca discrete modes $P_{prd}$ (in units of
Casimir's parallel plate result) as a function of the circumference
$d$ of the compact dimension (in units of the plate separation $a$)}
\end{center}
\end{figure}

\begin{figure}[ht]
\begin{center}
\includegraphics[scale=0.80]{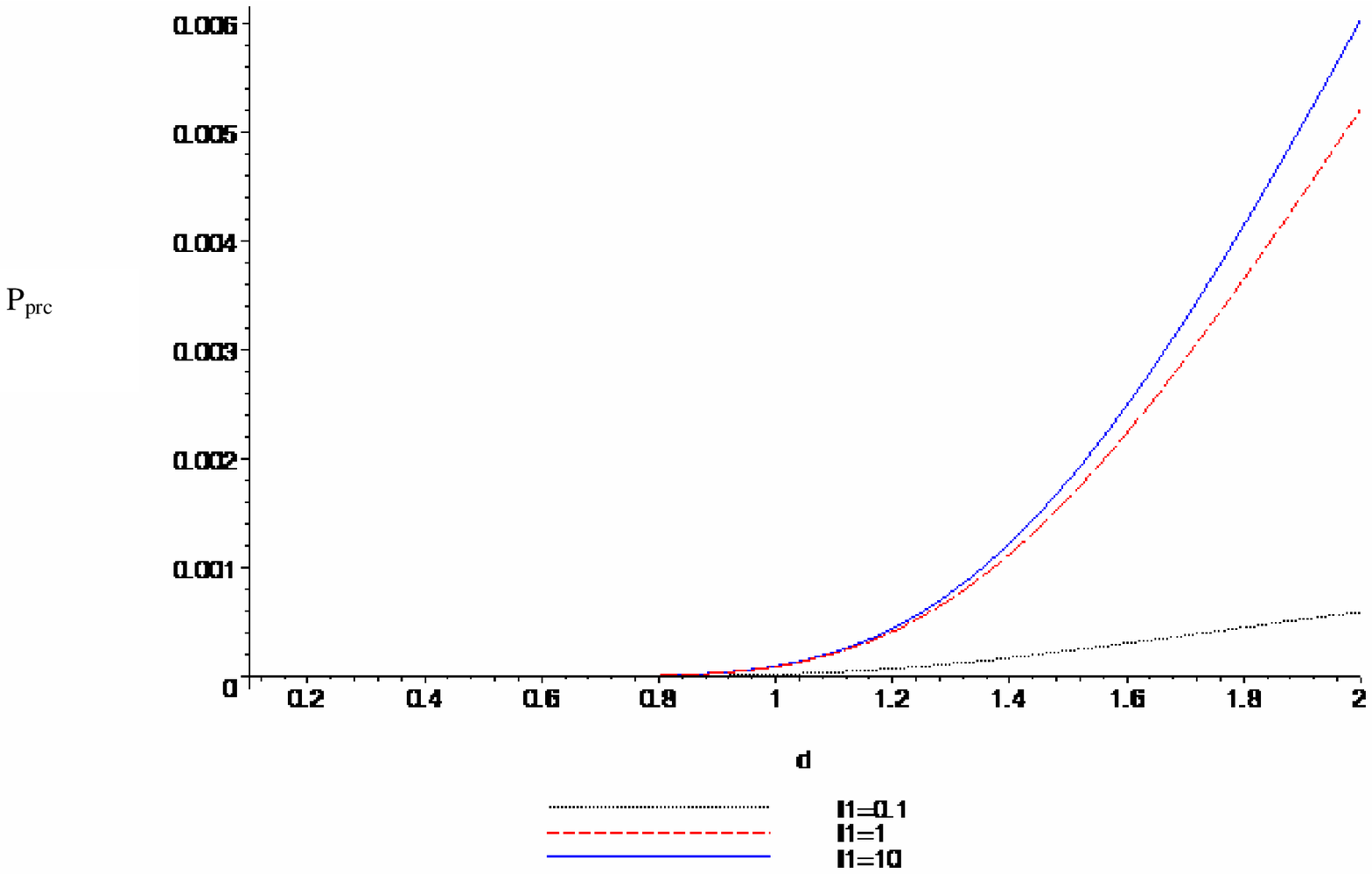}
\caption{Correction from the Proca continuum mode $P_{prc}$ (in
units of Casimir's parallel plate result) as a function of the
circumference $d$ of the compact dimension and the thickness
$\ell_1$ of the two slabs ($d$ and $\ell_1$ are in units of the
plate separation $a$).}
\end{center}
\end{figure}


\begin{appendix}

\section{Manifestly negative expression for the Dirichlet Casimir piston in $d+1$-dimensions}
\def\theequation{A.\arabic{equation}}
\setcounter{equation}{0}

Consider a $d$-dimensional rectangular resonator $R=[0,a]\times M$,
where $M$ is a $d-1$-dimensional Dirichlet resonator  with a
boundary $\partial M$. Eigenfrequencies of the resonator $R$ with
Dirichlet boundary conditions are determined by
\begin{align}
& \omega^2 =  \Bigl(\frac{\pi l}{a}\Bigr)^2 + \lambda_{k D}^2 , \:\:
l=1 \: .. +\infty , k=1 \: .. +\infty \\
& \Delta^{(d-1)} f_k (x,y) = - \lambda_{k D}^2 f_k (x,y) \\
&  f_k (x,y) |_{\partial M} = 0 ,
\end{align}
$\Delta^{(d-1)}$ is a $d-1$-dimensional Laplace operator. One can
write $\lambda_{kD}$ explicitly:
\begin{equation}
\lambda_{kD} = \sqrt{\sum_{i=1}^{d-1} \Bigl(\frac{\pi
k_i}{L_i}\Bigr)^2 } ,
\end{equation}
here $L_i$ are lengths of the sides of the resonator $M$, $k_i$ are
positive integers.

We adopt the zeta function regularization \cite{Elizalde, Kirsten}. The Casimir energy is
defined then as follows:
\begin{equation}
E = \sum \frac{\omega^{1-s}}{2} \Bigl|_{s=0} . \label{a2}
\end{equation}
This sum has to be evaluated for large positive values of $s$, an
analytical continuation to the value $s=0$ is being performed at the
end of calculations.

Alternatively one can define the Casimir energy via a zero
temperature one loop effective action $W$ \cite{Vassilevich} ($T_1$ is a time interval
here):
\begin{align}
W &=  E T_1  \\
E &= - \zeta^{\prime} (0) \label{a3} \\ \zeta(s) &=
\frac{1}{\Gamma(\frac{s}{2})} \int_{0}^{+\infty} dt\, t^{\frac{s}{2}
- 1} \sum_{\omega} \int_{-\infty}^{+\infty} \frac{d p}{2\pi} \exp
\biggl( - t \Bigl(\frac{a}{\pi}\Bigr)^2 \Bigl(\omega^2
+p^2\Bigr)\biggr) \label{a1}
\end{align}
After integration over $p$ in (\ref{a1}) one can verify that
definitions (\ref{a2}) and (\ref{a3}) coincide.

In every Casimir sum it is convenient to write:
\begin{equation}
\sum_{l=1}^{+\infty} \exp(- t l^2) = \frac{1}{2} \, \theta_3
\Bigl(0, \frac{t}{\pi}\Bigr) - \frac{1}{2} . \label{a4}
\end{equation}
For the first term on the right-hand side of (\ref{a4}) we use the
property of the theta function $\theta_3(0,x)$:
\begin{equation}
\theta_3 (0, x) = \frac{1}{\sqrt{x}} \, \theta_3
\Bigl(0,\frac{1}{x}\Bigr)
\end{equation}
and the value of the integral
\begin{equation}
\int_0^{+\infty} dt \, t^{\alpha-1} \exp \Bigl(-p\: t
-\frac{q}{t}\Bigr) = 2 \Bigl(\frac{q}{p}\Bigr)^{\frac{\alpha}{2}}
K_{\alpha} (2 \sqrt{p q} )
\end{equation}
expressed in terms of a modified Bessel function $K_{\alpha}(x)$ for
nonzero values of $n$ to rewrite the zeta function $\zeta(s)$ in the
form:
\begin{multline}
\zeta (s) = \sum_{\lambda_{kD}} \int_{-\infty}^{+\infty}
\frac{dp}{2\pi} \biggl[
\frac{\sqrt{\pi}\:\Gamma\bigl((s-1)/2\bigr)}{2 \:\Gamma(s/2)}
\Bigl(\frac{a\sqrt{\lambda_{kD}^2 + p^2}}{\pi}\Bigr)^{1-s}
 \\ + \sum_{l=1}^{+\infty} \frac{2\sqrt{\pi}}{\Gamma(s/2)}
\Bigl(\frac{\pi^2 l}{a \sqrt{\lambda_{kD}^2
+p^2}}\Bigr)^{\frac{s-1}{2}} K_{\frac{s-1}{2}} \Bigl(2al
\sqrt{\lambda_{kD}^2 +p^2}\Bigr) \biggr]   \\ +
\sum_{\lambda_{kD}}\frac{\sqrt{\pi}\:\Gamma\bigl((s-1)/2\bigr)}{4 a
\Gamma(s/2)} \Bigl(\frac{a \lambda_{kD}}{\pi}\Bigr)^{1-s} \label{a5}
\end{multline}

The Casimir energy of a resonator $R$ is given by:
\begin{multline}
E = \sum_{\lambda_{kD}} \int_{-\infty}^{+\infty} \frac{d p}{2 \pi}
\frac{1}{2} \ln \Bigl(1-
\exp(-2 a \sqrt{\lambda_{kD}^2 +p^2}) \Bigr)  \\
+ \frac{a}{2} \sum_{\lambda_{kD}} \int_{-\infty}^{+\infty} \frac{d
p}{2 \pi}  \Bigl( \lambda_{kD}^2 +p^2
\Bigr)^{\frac{1-s}{2}}\biggr|_{s=0} + \frac{1}{4}
\sum_{\lambda_{kD}} \lambda_{kD}^{1-s}\biggr|_{s=0} . \label{a6}
\end{multline}
Here we used the property $K_{-1/2} (x) =\sqrt{\pi/(2\,x)}\exp(-x)$.

The term
\begin{equation}
E_{cylinder} = \frac{1}{2} \sum_{\lambda_{kD}}
\int_{-\infty}^{+\infty} \frac{d p}{2 \pi} \Bigl( \lambda_{kD}^2
+p^2 \Bigr)^{\frac{1-s}{2}}\biggr|_{s=0} \label{om8}
\end{equation}
can be thought of as the Casimir energy per unit length of an
infinite cylinder with Dirichlet boundary conditions and the same
$d-1$-dimensional section $M$ as the resonator $R$.

For the experimental check of the Casimir energy one should measure
the force. One can insert two $d-1$-dimensional plates $M$ inside an
infinite cylinder with the same $d-1$-dimensional cross section $M$
and measure the force acting on one of the plates as it is being
moved through the cylinder.  The distance between the plates is $a$.


To calculate the force on each of the two plates inside a cylinder
with the cross section $M$ one can perform the following gedanken
experiment that was frequently used to calculate the Casimir force
between two infinite parallel plates. Imagine that $4$ parallel
plates are inserted inside an infinite cylinder and then $2$
exterior plates are moved to spatial infinity. This situation is
exactly equivalent to $3$ cavities touching each other. From the
energy of this system one has to subtract the Casimir energy of an
infinite cylinder without plates inside it, only then do we obtain
the energy of interaction between the interior parallel plates, the
one that can be measured in the experiment. The force on each of the
interior parallel plates is the same as the force on the piston when
one of the three piston plates (the exterior plate) is moved to
infinity. So the attractive force on each of the $d-1$-dimensional
Dirichlet pistons inside the $d$-dimensional Dirichlet cylinder is
equal to:
\begin{equation}
F (a) = - \frac{\partial \mathcal{E}(a)}{\partial a}, \label{a13}
\end{equation}
where the Casimir energy of a $d-1$-dimensional piston with
Dirichlet boundary conditions can be written as follows:
\begin{align}
 \mathcal{E} (a) &= \sum_{\omega_c} \frac{1}{2} \ln (1-\exp(-2
a \, \omega_c)) = \nonumber \\ &=\sum_{\lambda_{kD}}
\int_{-\infty}^{+\infty} \frac{d p}{2 \pi} \frac{1}{2} \ln \Bigl(1-
\exp(-2 a \sqrt{\lambda_{kD}^2 +p^2}) \Bigr) , \label{r7}
\end{align}
the sum here is over all  eigenfrequencies $\omega_c$ for a cylinder
with a $d-1$-dimensional cross section $M$ and an infinite length
with Dirichlet boundary conditions imposed.

By making use of an identity \cite{Nesterenko}
\begin{equation}
\frac{1}{2} \int_{-\infty}^{+\infty} \frac{dp}{2\pi} \ln
\bigl(1-\exp(-2a\sqrt{\lambda^2 + p^2})\bigr) = -
\frac{\lambda}{2\pi} \sum_{l=1}^{+\infty} \frac{K_1(2l\lambda\,
a)}{l}
\end{equation}
 one can rewrite (\ref{r7}) in the form:
 \begin{align}
 &\mathcal{E}(a) = -\frac{1}{2\pi} \sum_{l=1}^{+\infty} \sum_{\lambda_{k D}}
 \frac{\lambda_{k D} K_1 (2 l\lambda_{k D} a)}{l}  , \\
 &\lambda_{kD} = \sqrt{\sum_{i=1}^{d-1} \Bigl(\frac{\pi
k_i}{L_i}\Bigr)^2 } \nonumber \label{p10}
 \end{align}
the sum should be performed over all sets of positive integer values
of $k_i$.

\end{appendix}
\section*{Acknowledgments}

AE acknowledges support from a discovery grant of the Natural
Sciences and Engineering Research Council of Canada (NSERC). VM
acknowledges support from a CNRS grant ANR-06-NANO-062 and grants
RNP 2.1.1.1112, SS 5538.2006.2 and RFBR 07-01-00692-a. AE would like
to thank the Kavli Institute for Theoretical Physics (KITP, Santa
Barbara) for their hospitality and financial support where part of
this work was completed and presented (this research was supported
in part by the National Science Foundation under Grant No.
PHY05-51164). AE would also like to thank professor Raman Sundrum of
John Hopkins University for email correspondence.


\begin{thebibliography}{99]}
\bibitem{Kaluza} T. Kaluza, Preus. Acad. Wiss. K1, 966 (1921) ;
\bibitem{Klein} O. Klein, Z. Phys. {\bf 37},  895 (1926).
\bibitem{Greene} B.R. Greene and J. Levin, JHEP {\bf 096} (2007) 0711
[arXiv:0707.1062].
\bibitem{Setare} A.A. Saharian and M.R. Setare, Phys. Lett. B {\bf 659}, 367
(2008) [arXiv:0707.3240]; A.A. Saharian and M.R. Setare, Phys. Lett.
B {\bf 552}, 119 (2003) [hep-th/0207138].
\bibitem{Mariana-Frank} M. Frank, N. Saad and I. Turin,
[arXiv:0807.0443].
\bibitem{Cleaver} R. Obousy and G. Cleaver, [arXiv:0810.1096].
\bibitem{Cheng} H. Cheng, arXiv:0801.2810.
\bibitem{Leandros} L. Perivolaropoulos, Phys. Rev. D {\bf 77}, 107301
(2008) [arXiv:0802.1531].
\bibitem{Bleicher} K. Poppenhaeger, S. Hossenfelder, S. Hofmann and
M. Bleicher, Phys. Lett. B {\bf 582}, 1 (2004).
\bibitem{Raman} R. Sundrum, TASI lectures hep-th/0508134.
\bibitem{Dombey} G. Barton and N. Dombey, Nature {\bf 311}, 336, (1984);
G. Barton and N. Dombey, Ann. of Phys. {\bf 162}, 231 (1985).
\bibitem{Lifshitz} E.~M. Lifshitz, Zh.Eksp.Teor.Fiz. {\bf 29}, 94 (1955) ;  Soviet Phys. JETP {\bf 2}, 73 (1956).
\bibitem{Casimir} H. B. G. Casimir, Proc. Kon. N. Akad. Wet. {\bf 51}, 793 (1948).
\bibitem{Marachevsky1} V.N. Marachevsky, Phys. Rev. D {\bf 75}, 085019 (2007) [hep-th/0703158].
\bibitem{Cavalcanti} R. M. Cavalcanti, Phys. Rev. D {\bf 69}, 065015
(2004).
\bibitem{Ariel1} A. Edery, Phys. Rev. D {\bf 75}, 105012 (2007)[hep-th/0610173];
J. Phys. A: Math. Gen. {\bf 39}, 685 (2006)[math-ph/0510056].

\bibitem{Hertzberg}
M.P. Hertzberg, R.L. Jaffe, M. Kardar and A. Scardicchio, Phys. Rev.
Lett.{\bf 95}, 250402 (2005) [quant-ph/0509071];  Phys. Rev. D {\bf
76}, 045016 (2007) [arXiv:0705.0139].
\bibitem{ArielVal} A. Edery and V. Marachevsky, Phys. Rev. D {\bf 78 }, 025021 (2008).
[arXiv:0805.4038].
\bibitem{Marachevsky1A} V.N. Marachevsky, J. Phys. A: Math. Theor. {\bf 41}, 164007 (2008).
[arXiv:0710.4130].
\bibitem{Ariel3} A. Edery and I. MacDonald, JHEP {\bf 09}, 005  (2007) [arXiv:0708.0392].
\bibitem{Marachevsky2} V.N. Marachevsky, hep-th/0609116, published in the
proceedings of QUARKS-2006.
\bibitem{Marachevsky3} V.N. Marachevsky, hep-th/0512221.
\bibitem{Barton} G. Barton, Phys. Rev. D {\bf 73}, 065018 (2006).
\bibitem{Ariel2} A. Edery, J. Stat. Mech. P06007 (2006) [hep-th/0510238].
\bibitem{Fulling} S.A. Fulling, L. Kaplan and J.H. Wilson, Phys. Rev. A {\bf 76}, 012118 (2007) [arXiv:0703248].
\bibitem{Lim1} S. C. Lim and L.P. Teo,  [arXiv:0807.3613].
\bibitem{Lim2} S. C. Lim and L.P. Teo,  [arXiv:0808.0047].
\bibitem{Ginzburg} Yu.S. Barash and V.L. Ginzburg, Sov.Phys.Usp. {\bf 18},
305 (1975).

\bibitem{Milton1}
K.A. Milton, J.Phys.A: Math. Gen. {\bf 37}, R209 (2004).

\bibitem{Brevik1}
S.A. Ellingsen and I. Brevik, J.Phys.A: Math.Gen. {\bf 40}, 3643
(2007).

\bibitem{gratings} A. Lambrecht and V.N. Marachevsky, Phys. Rev. Lett. {\bf 101}, 160403 (2008) [arXiv:
0806.3142].
\bibitem{Davies} P.C.W. Davies and S. Unwin, Phys. Lett. B {\bf 98}
274, (1981).
\bibitem{Onofrio} G. Bressi, G. Carugno, R. Onofrio and G. Ruoso,
Phys. Rev. Lett. {\bf 88}, 041804 (2002).



\bibitem{Mohideen} U. Mohideen and A. Roy, Phys. Rev. Lett. {\bf 81},4549
(1998);  A. Roy, C.Lin and U. Mohideen, Phys. Rev. D {\bf 60},111101
(1999); M.Bordag, U.Mohideen and V.M.Mostepanenko, Phys.Rep. {\bf
353}, 1 (2001).

\bibitem{Decca}
R.S. Decca, D. Lopez, E. Fischbach, G.L. Klimchitskaya, D.E. Krause,
V.M. Mostepanenko,  Ann.Phys. {\bf 318}, 37 (2005); Eur.Phys.J.C
{\bf 51}, 963 (2007).

\bibitem{Elizalde}
E.Elizalde, {\it Ten physical applications of spectral zeta functions} (Springer)  (1995).

\bibitem{Kirsten}
K.Kirsten, {\it Spectral functions in Mathematics and Physics} (Boca Raton, FL: CRC Press) (2002).

\bibitem{Vassilevich}
D.V. Vassilevich, Phys.Rep. {\bf 388},  279  (2003).

\bibitem{Nesterenko}
V.V. Nesterenko and I.G. Pirozhenko, J.Math.Phys.{\bf 38}, 6265
(1997) [arXiv: hep-th/9703097].

\end{thebibliography}
\end{document}